\documentclass[12pt]{article}

\usepackage{amsmath}
\usepackage{amssymb}
\usepackage[linktocpage=true, colorlinks=false, linkcolor = blue, citecolor = red]{hyperref}
\usepackage{xfrac}
\usepackage{pgfplots}
\usepackage{cite}
\usepackage{tikz}
\usepackage{xcolor}

\newcommand{\ls}{\left(}
\newcommand{\rs}{\right)}
\newcommand{\ta}{\tau}
\newcommand{\ff}{\varphi}
\newcommand{\te}{\theta}
\newcommand{\sign}{{\rm sign}}
\newcommand{\sh}{\sinh}
\newcommand{\ch}{\cosh}
\newcommand{\ctg}{\cot}

\newcommand{\disn}[2]{$$\displaylines{\refstepcounter{equation}%
            \label{#1}\hskip 1em minus 1em #2\hfilneg}$$}
\newcommand{\nom}{\hfil\hskip 1em minus 1em (\theequation)}
\newcommand{\no}{\hfil \hskip 1em minus 1em\phantom{(\theequation)}%
            \hfilneg\cr\hfilneg\hskip 1em minus 1em\hfil}
\newcommand{\ns}{\hfill\cr\hfill}

\newcommand{\bq}{\begin{equation}}
\newcommand{\eq}{\end{equation}}

\begin{document}

\title{Explicit isometric embeddings\\ of collapsing dust ball}
\author{A.~D.~Kapustin, M.~V.~Ioffe, S.~A.~Paston\thanks{E-mail: pastonsergey@gmail.com}\\
{\it Saint Petersburg State University, Saint Petersburg, Russia}
}
\date{\vskip 15mm}
\maketitle

\begin{abstract}
The work is devoted to the search for explicit isometric embeddings of a metric corresponding to the collapse of spherically symmetric matter with the formation of a black hole.
Two approaches are considered: in the first, the embedding is constructed for the whole manifold at once; in the second, the idea of a junction of solutions, obtained separately for areas inside and outside the dust ball, is used.
In the framework of the first approach, a global smooth embedding in 7D space with a signature (2 + 5) was constructed. It corresponds to the formation of the horizon as a result of matter falling from infinity.
The second approach generally leads to an embedding in 7D space with the signature (1 + 6). This embedding
corresponds to the case when matter flies out of a white hole with the disappearance of its horizon,
after which the radius of the dust ball reaches its maximum, and then a collapse occurs with the formation of the horizon of a black hole.
The embedding obtained is not smooth everywhere --- it contains a kink on the edge of the dust ball, and {also, it is} not quite global.
In the particular case, when the maximum radius of the dust ball coincides with the radius of the horizon, it is possible to construct a global smooth embedding in a flat 6D space with a signature (1 + 5).
\end{abstract}


\newpage

\section{Introduction}
It is known (see for example \cite{goenner}), that an arbitrary (pseudo)Riemannian $d$-dimensional manifold could be \emph{isometrically} embedded in a \emph{flat} ambient space of dimension $N \geqslant d(d+1)/2$ at least locally.
As a result a manifold can be {described} using embedding {function} $y^a(x^\mu)$, and a metric can be considered as the \emph{induced} one
\bq\label{s1}
g_{\mu \nu} = (\partial_{\mu} y^a) (\partial_{\nu} y^b) \eta_{ab},
\eq
where $\eta_{ab}$ is a flat ambient space metric; hereafter $\mu,\nu=0,...,d-1$; $a,b=1,...,N$.
Such definition of manifold could be visual and useful for its structure study, but it requires an explicit expression for embedding {function} of given metric $g_{\mu\nu}$, i.e. one has to solve the differential equation \eqref{s1} w.r.t. $y^a$. The manifold structure study is especially {important} in the case of black holes because {the} corresponding manifolds usually have non-trivial structure.

The first explicit embedding was constructed in 1921 for the Schwarzschild metric corresponding to the non-rotating, uncharged {black hole \cite{kasner3}.
However, it (and also, embedding \cite{fudjitani}) covers} only the region outside of the horizon
{and hence, it is not suitable for} studying the global structure of the manifold.
The most useful embedding for such purpose was proposed in 1959 in the work \cite{frons}. It is smooth everywhere
and covers both inside and outside regions of the horizon. In addition, this embedding corresponds to the maximal analytic  extension of the Schwarzschild solution,
which includes two areas outside of the horizon (two universes) and two areas inside the horizon, corresponding to a white hole and a black hole.
Beside these embeddings, there are another three so-called
{"minimal"{}, i.e. with minimal possible dimensionality of the ambient space
(6 in that case, {see \cite{kasner2}}), embeddings {\cite{davidson,statja27}}
of the Schwarzschild metric. They cover a half} of the already mentioned maximal analytic extension --- e.g., a set of the area outside and area inside,
corresponding to the black hole {(see \cite{statja27} for details).}
{Note, that} only these two areas exist in the case of the Riemannian manifold
{if the black hole results from the collapse, while the maximal analytic extension corresponds to the {\it eternal} black hole.}

The problem of finding the minimal \emph{global} (i.e. smooth for any radius value, including horizon points)
embeddings for the non-rotating Reissner-Nordstrom black hole {was} studied in the work \cite{statja30}.
Three variants of embeddings were constructed,
{which can be used both in the case of non-extremal charged black hole, and in the cases of extremal and hyperextremal one.}
A generalization to the case of a non-zero cosmological constant was studied in the work \cite{statja40}.
In a more physically
{interesting cases of the rotating Kerr black hole and its generalization -- the charged rotating Kerr-Newman black hole, the problem}
of constructing an explicit embedding becomes much more
{complicated due to smaller symmetry of the {metric}.}
{Currently, only two {of embedding options} are known for
such black holes. The first is a
local embeddings in the 9-dimensional ambient space [10] and [11]
(for the Kerr and Kerr-Newman metric, respectively) when
2 out of 9 components of the embedding {function} are described as solutions of
some ODE, i.e. implicitly. And the second is a
14-dimensional embedding of the Kerr metric proposed in [12].}

The construction of explicit embeddings for physically interesting solutions of {General Relativity} could be useful
{in studying the embedding theory --- the Regge-Teitelboim alternative gravity theory,} originally proposed in the work \cite{regge}.
Within this approach, {just the embedding function $y^a(x)$ is} an independent variable instead of a metric
{which is defined by} formula \eqref{s1}.
After \cite{regge}, an idea of isometric embedding as a tool for description of gravity (and for its quantization, as well) has  regularly been discussed by many authors (see, e.g., papers \cite{deser,pavsic85let,tapia,davkar,statja18,rojas09,statja25,faddeev,statja51}).
One can find a detailed list of
{references on the embedding theory and close issues in the review} \cite{tapiaob}.

The explicit embeddings of Riemannian manifolds with horizon {are} also used in the analysis of a connection between Hawking radiation and Unruh radiation corresponding to the movement of the observer in the ambient space \cite{deserlev98,deserlev99,statja34,statja36},
{{(see also references in \cite{statja34})}.
Using this connection the studies of the black holes thermodynamics is developing
\cite{Dunajski2019,Govindarajan2019,1905.04860}.
The idea of embedding a metric in the flat space of a larger number of dimensions also continues to be successfully used for search an exact solution of the Einstein-Maxwell field equations, corresponding to stellar structures
\cite{Bhar2017,Maurya2017-2},
including the framework of modified gravity theories \cite{Maurya2019}.}

The main difficulty in constructing explicit embeddings for arbitrary 4-dimensional space-time is that we need to solve a system of 10 PDEs \eqref{s1} w.r.t embedding {function} $y^a(x^\mu)$,
depending on $4$ coordinates $x^\mu$.
The problem is simplified for the manifolds with additional symmetries:
{for the symmetry group large enough, we can}
use the constructive method of finding explicit embeddings \cite{statja27}
which could reduce the system \eqref{s1} to the system of ODEs.
This is exactly what happens for the Schwarzschild and Reissner-Nordstrom metrics, with the symmetry group $SO(3)\times T^1$, where $T^1$ denotes the group of time translations. A similar situation occurs (see~\cite{statja29}) in case of cosmological solutions --- for the metrics of all three FRW models, with symmetries: $SO(4)$ for the closed model, $SO(1,3)$ for the open model and a group of movements of the three-dimensional plane for the spatially-flat model.

Probably the most physically interesting variant of a black hole is a black hole arising from a collapse when a cloud of matter shrinks and a black hole is generated dynamically. In such a process, a horizon formation occurs and hence a study of the structure of corresponding manifold becomes very interesting, so the problem of constructing an explicit embedding becomes relevant in that case.
Even if we neglect rotation, i.e. if we will consider that the metric corresponds to Schwarzschild solution in the space around the matter cloud, the problem of constructing an explicit embedding for that metric seems to be very difficult, and up to now, it has not been done. {Just the problem} of the construction of such embeddings is considered in the present paper. To simplify the problem, we take the simplest variant of the collapsing matter behavior and consider the collapse of a homogeneous ball consisting of a dust-like matter.

The symmetry group of this problem is $SO(3)$, and this symmetry is not large enough to reduce the problem to the solution of the ODE system in the framework of the method \cite{statja27}. However, if we consider a manifold as a set of two parts --- the first contains matter (compressing or expanding dust ball) and the second doesn't contain it (area outside that ball), then a symmetry group is larger for each part and it can simplify the problem.
{According to the well-known Birkhoff theorem, the metric outside the ball is the}
Schwarzschild metric. Therefore, it has the symmetry $SO(3) \times T^1$ and we know for it the variety of embeddings mentioned above.
{Inside the ball, the metric corresponds to one of the FRW models (e.g., see \cite{landavshic2}). Hence,}
it also has an extended symmetry of the corresponding type and the embeddings for this metric are also known \cite{robertson1933}.

The junction of appropriately modified known embeddings of two parts will lead us to the embedding for the whole manifold. This method is used in the Section~4. For the more interesting case of a dynamically generated horizon we have succeeded in constructing an embedding in a 7-dimensional ambient space with signature $(+------)$. However, it contains a kink (discontinuity of the first derivative of the embedding function), and also it cannot be extended to an area of arbitrarily large radius. For the case of the static horizon with the matter completely under it (flying out of a white hole and falling into a black hole), a smooth embedding into 6-dimensional ambient space with signature $(+-----)$ is constructed.

An alternative way is to construct the desired embedding {in a single way,} without using a junction of known embeddings. In this case, homogeneity of a dust ball simplifies the problem and allows us to find an explicit embedding for collapse, i.e. for the case of a dynamically generated horizon.
This approach is used in Section~3.
In this case, it is possible to construct an embedding in a 7-dimensional ambient space with the signature $(+-+----)$, and it turns out to be smooth.

{Obtained explicit embeddings could be useful for visualization of the geometry corresponding to the process of collapse.
{Especially, it seems to be interesting in the case of dynamically formation of the horizon.
These embeddings can be also used to study the thermodynamics of a black hole.
{At that it} is important to keep in mind that for obtained embeddings {structure}, the exact}
Hawking into Unruh mapping will not be performed \cite{statja36}.}

\section{Expression for the metric and used coordinate frames}
We write the expression for the {spherically-symmetric} metric that corresponds to a compressing (or expanding) dust ball of finite size. This metric should be a solution {of} the Einstein equations
\begin{equation}
\label{G=T}
G_{\mu \nu} = \varkappa\, T_{\mu \nu}
\end{equation}
with energy-momentum tensor (EMT) corresponding to the {mentioned} kind of matter and its distribution.
The dustlike EMT has the simplest form if we use a {synchronous} comoving coordinate frame.
Due to the spherical symmetry, it is convenient to use angles $\theta$ and $\varphi$ as two spatial coordinates.
The remaining timelike coordinate we denote as $\tau$, and the spatial one as $\chi$.

{Such} solution of the Einstein equations can be found in the form of a diagonal metric. For the arbitrary distribution of {dustlike} matter w.r.t. radius a corresponding squared interval (further we will refer to such formulas as metric) has the form \cite{landavshic2}
\bq
\label{metric}
d s^2 = d \tau^2 - \frac{(r'(\tau, \chi))^2}{1+f(\chi)} d\chi^2 - r^2(\tau, \chi) d \Omega^2,
\eq
where $d\Omega^2=d\theta^2+\sin^2\theta d\varphi^2$, and
function $r(\tau, \chi)$ is determined by one of three ways
\bq
\label{v123}
r(\tau, \chi) = \left\{
\begin{array}{l}
\vphantom{\Biggr(}
\dfrac{F(\chi)}{2 f(\chi)} H \left( \dfrac{2(f(\chi))^{\sfrac{3}{2}} }{F(\chi)}(\tau_0(\chi)-\tau) \right) , \ \ \ \ f(\chi) > 0,\\
\left( \dfrac{9F(\chi)}{4} \right)^{\sfrac{1}{3}} \left[ \tau_0(\chi)- \tau \right]^{\sfrac{2}{3}}, \ \ \ \ f(\chi) = 0, \\
- \dfrac{F(\chi)}{2 f(\chi)} E \left( \dfrac{2(-f(\chi))^{\sfrac{3}{2}} }{F(\chi)}(\tau_0(\chi)-\tau) \right), \ \ \ \ f(\chi) < 0.
\end{array} \right.
\eq
Functions $H(x)$ and $E(x)$ {are} used for the inversion of parametric {dependence}
\bq\label{sp4}
H = \ch{\eta}-1,\quad x = \sh{\eta} - \eta;\qquad
E = 1 - \cos{\eta},\quad x = \eta - \sin{\eta},
\eq
and functions $F(\chi), f(\chi)$ and $\tau_0(\chi)$ {all} {together} define {the distribution of a matter density and its initial speed}.
{{At that,}}
the function $F(\chi)$ has the meaning of the Schwarzschild radius of {the} matter {with} coordinate {values} less than $\chi$, {while} the function $\tau_0 (\chi)$ determines the \textit{moment of} time $\tau$ at which the particle with coordinate $\chi$ reaches the point $r = 0$, i.e. {it falls} on the singularity {produced by} the collapse. The function $f(\chi)$ has no {specific} well-defined physical sense, however, depending on it's sign function $f(\chi)$ classifies the space-time as bound, marginally bound
or unbound if $f(\chi)$ is negative, null or positive, respectively, see \cite{joshi} for more details.
If we substitute the metric \eqref{metric} into the Einstein equation \eqref{G=T}, we obtain the expression for EMT:
\begin{equation}
\label{EMT}
	T_{\mu \nu} = \rho(\tau, \chi) \, \delta_\mu^0 \, \delta_\nu^0, \qquad \rho(\tau, \chi) = \frac{F'(\chi)}{\varkappa\, r^2(\tau, \chi) \,  r'(\tau, \chi)},
\end{equation}
where prime denote derivatives with respect to variable $\chi$. These formulas can be also found in \cite{joshi}.

{For the homogeneous initial distribution of matter, the space inside the collapsing ball must be described by the geometry of the open, spatially flat or closed FRW model, {respectively}, by choosing the first, second or third method of determining $ r(\tau, \chi)$ in the formula \eqref{v123}.}
It is easy to notice that the homogeneity at each moment of time requires the simultaneity of the fall of all particles to the point $r = 0$, {and therefore,} the function $\tau_0(\chi)$ should {become} a constant {which can be taken zero by choosing the time $\tau$.}

According to the Birkhoff theorem, the space outside of the ball is always described by the Schwarzschild geometry.
As we can see from \eqref{EMT} in this area $F (\chi)$ should be constant, since in that case EMT vanishes.
Therefore, the substitution of $F (\chi) = const$ into (\ref{metric})
leads to the Schwarzschild metric in some coordinate frame.

If a matter density inside the ball is constant, and hence, has a jump on the boundary of the ball, {the} metric (\ref{metric}) will have {the corresponding jump-like coordinate singularity}, which can be avoided by the change of the coordinate frame to the $(\tau, r, \theta, \varphi)$.
{After this change of coordinates, the} metric \eqref{metric} takes the form
\bq\label{sp1}
	d s^2 = \left( 1- \frac{\dot{r}(\tau, \chi)^2}{1+f(\chi)}\right) d \tau^2 + 2\frac{\dot{r}(\tau, \chi)}{1+f(\chi)}dr d\tau - \frac{dr^2}{1+f(\chi)} - r^2 d\Omega^2,
\eq
where $\dot{r}(\tau, \chi) \equiv \frac{\partial r(\tau, \chi)}{\partial \tau}$, and $r(\tau, \chi)$ is given by \eqref{v123}.
The value $\chi$ entering into \eqref{sp1} through $\dot{r}(\tau, \chi)$ and $f(\chi) $ must be expressed in terms of the independent coordinates $ \tau $ and $ r $ according to \eqref{v123}.

{Now, if we choose the option $f(\chi)=0$ in \eqref{v123}, the situation will become much simpler,} because $r(\tau, \chi)$ and $\dot{r}(\tau, \chi)$ can be written explicitly. After simple transformations, the metric will take the form
\bq
\label{metric2}
	d s^2 = \left(1-\frac{F(\tau, r)}{r} \right)d\tau^2 - 2 \sqrt{\frac{F(\tau, r)}{r}}dr d\tau  - dr^2 - r^2 d\Omega^2,
\eq
where hereinafter $F(\tau, r)\equiv F(\chi(\tau, r))$.
While the matter density remains finite, this metric, written in the coordinates $(\tau, r, \theta, \varphi)$, will be continuous due to the continuity of the function $F(\chi)$ in this case.

Inside the homogeneous dust ball, we can assume $\tau_0(\chi)=0$ (see {comment after} \eqref{sp4}), which {allows to find {in this area} the expression for function $F(\tau, r)$ from \eqref{v123}, in the case $f(\chi)=0$}. {Taking into account that $F(\chi) = const$ outside the ball and assuming continuity of $F(\chi)$, one can conclude:}
\bq\label{sp2}
	F(\tau, r) = \min{ \left( \frac{4r^3}{9\tau^2}, F_0 \right) },
\eq
where $F(\tau, r)=F_0$ outside of the ball. Here, we assume that the time $ \tau $ is everywhere {negative, with $\tau = 0$ corresponding to the end of the collapse,} when the whole matter simultaneously reached the point $r=0$ and formed the singularity.

The expression for the metric \eqref{metric2} in coordinates $(\tau, r, \theta, \varphi)$ will be {used to build} an explicit embedding in Section~3.
The possibility to write the metric in the form \eqref{metric} in the synchronous and co-moving coordinate frame will be used in Section~4.

\section{Constructing the embedding without using a junction of solutions}
We will construct the first variant of an explicit embedding for metric (\ref{metric2}) {without
joining the} separate embeddings for the {areas} inside and outside of the ball.
In this case, however, we will assume that the dust ball is homogeneous at each moment and the formula \eqref{sp2} can be used
{in this case to simplify the problem.}
Recall that the value of $F(\tau, r)$ given by this formula, up to a factor, gives the mass of dust contained within the radius $r$ at the time $\tau$. {
According} to \eqref{sp2} the metric (\ref{metric2}) is continuous, {but its} first derivative has a jump on the boundary of the ball.

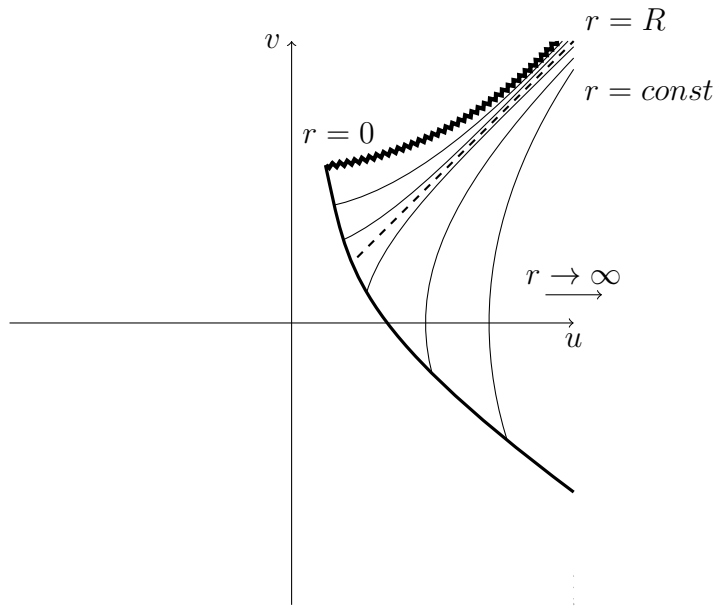
\begin{figure}[h!]
	\centering
	\begin{tikzpicture}[scale=0.75]
		\draw[->] (-5, 0) -- (5, 0)node[below]{$u$};
		\draw[->] (0, -5) -- (0, 5)node[left]{$v$};
		\draw[->] (4.5, 0.5) -- (5.5, 0.5)node[midway,above]{$r \to \infty$};
		\begin{scope}
		\clip (7.5, 6) -- (5, 6) -- (0.6, 2.8) .. controls (1, 1) and (1, 0) .. (5, -3) -- (5, -6) -- (7.5, -6) -- cycle;
		\draw[thick, dashed] (-5, -5) -- (5, 5)node[above right]{$r = R$};
		\draw[thick, dashed] (-5, 5) -- (5, -5);
		\draw (5, 4.9) .. controls (0, 0) and (0, -0) .. (5, -4.9);
		\draw (5, 4.7) .. controls (1.5, 1) and (1.5, -1) .. (5, -4.7);
		\draw (5, 4.5)node[below right]{$r=const$} .. controls (3, 1.5) and (3, -1.5) .. (5, -4.5);
		\draw (-5, 4.9) .. controls (0, 0) and (0, -0) .. (-5, -4.9);
		\draw (-5, 4.7) .. controls (-1.5, 1) and (-1.5, -1) .. (-5, -4.7);
		\draw (-5, 4.5) .. controls (-3, 1.5) and (-3, -1.5) .. (-5, -4.5);
		\draw (4.9, 5) .. controls (0, 0) and (0, 0) .. (-4.9, 5);
		\draw (4.8, 5) .. controls (1, 1) and (-1, 1) .. (-4.8, 5);
		\draw[ultra thick, decoration = {zigzag,segment length = 1mm, amplitude = 0.3mm}, decorate] (4.7, 5) .. controls (2, 2) and (-2, 2) .. (-4.7, 5);
		\draw (4.9, -5) .. controls (0, 0) and (0, 0) .. (-4.9, -5);
		\draw (4.8, -5) .. controls (1, -1) and (-1, -1) .. (-4.8, -5);
		\draw[ultra thick, decoration = {zigzag,segment length = 1mm, amplitude = 0.3mm}, decorate] (4.7, -5) .. controls (2, -2) and (-2, -2) .. (-4.7, -5);
		\draw (0, -3)node[below right]{$r = 0$};
		\end{scope}
		\draw (0, 3)node[above right]{$r = 0$};
		\draw[very thick] (5, -3) .. controls (1, 0) and (1, 1) .. (0.6, 2.8);
	\end{tikzpicture}
	\caption{\label{infkrusk}The Kruskal diagram for the matter collapsing from infinity {($u,v$ {are} the Kruskal-Szekeres coordinates).}}
\end{figure}
{The area outside of the ball, described by the Schwarzschild metric, is shown on the Kruskal diagram in Fig.~\ref{infkrusk}.
Kruskal-Szekeres coordinates \cite{kruskal} {are} defined as
\begin{equation}
\begin{aligned}
        &\qquad\qquad\qquad r > R:\\
		&\left\{ \begin{aligned}
			u &= \left( \frac{r}{R}-1 \right)^{\sfrac{1}{2}} e^{\frac{r}{2R}} \ch{\left( \frac{t}{2R} \right)} \\
			v &= \left( \frac{r}{R}-1 \right)^{\sfrac{1}{2}} e^{\frac{r}{2R}} \sh{\left( \frac{t}{2R} \right)}
		\end{aligned} \right.,
\end{aligned}
		\qquad
\begin{aligned}
        &\qquad\qquad\qquad r < R:\\
		&\left\{ \begin{aligned}
			u &= \left( 1-\frac{r}{R} \right)^{\sfrac{1}{2}} e^{\frac{r}{2R}} \sh{\left( \frac{t}{2R} \right)} \\
			v &= \left( 1-\frac{r}{R} \right)^{\sfrac{1}{2}} e^{\frac{r}{2R}} \ch{\left( \frac{t}{2R} \right)}
		\end{aligned} \right..
\end{aligned}
\end{equation}
The area outside of the ball
is limited by the world line of the point corresponding to the boundary of the ball.}
The remaining part of the manifold, corresponding to the interior of the ball, is described by the FRW metric and is not shown in the figure. Since we take $f(\chi)=0$ when moving from (\ref{metric}) to (\ref{metric2}), this is a metric of the spatially flat FRW model (see \eqref{v123} and {the text after it).}

It can be seen from \eqref{sp2} that the boundary of the ball is described by the equation
\disn{sp3}{
r^3=\frac{9}{4}F_0\tau^2,
\nom}
{so that}
$r\to\infty$ when $\ta\to-\infty$,
i.e. the size of the dust ball was infinite in the past.
It means that the embedding, constructing in this section, corresponds to the collapse of particles, falling from {arbitrary large distance at the initial moment}. Note that this situation is qualitatively different from that considered in Section~4.

We will look for the embedding function $y^a(x^\mu)$ {as a set} of components $\{\tilde y^A(\tau,r),$ $\hat y^i(r,\te,\ff)\}$, where three components $\hat y^i(r,\te,\ff)$ have a usual form for the spherically symmetric embeddings\begin{align}\label{sp7}
\hat y^1 &= r \cos{\theta}, \nonumber\\
\hat y^2 &= r \sin{\theta} \cos{\varphi},  \\
\hat y^3 &= r \sin{\theta} \sin{\varphi}.\nonumber
\end{align}
Then, taking into account the structure of the formula of the induced metric \eqref{s1}, the problem of constructing an explicit embedding for the metric (\ref{metric2}) is reduced to the problem of finding an embedding for a 2-dimensional surface
$\tilde y^A(\tau,r)$ with the metric
\disn{sp8}{
d s^2 = \left(1-\frac{F(\tau, r)}{r} \right)d\tau^2 - 2 \sqrt{\frac{F(\tau, r)}{r}}dr d\tau.
\nom}
{In the process of solving this problem, it is convenient to} introduce new variables $t = \tau^{\sfrac{2}{3}}$ and $p = r^3/\tau^2$ instead of $\tau$ and $r$, so \eqref{sp2} will have the form
\bq\label{sp6}
F(\tau, r) = \min{ \left( \frac{4}{9}p, F_0 \right) }\equiv\bar F(p).
\eq
Then we can obtain a following expression for the metric (\ref{sp8})
\bq\label{eq10}
	d s^2 = \left(\frac{9}{4} t +f(p) \right) dt^2 + t\, p^{-\sfrac{5}{6}}\sqrt{\bar F(p)} \, dp\, dt,
\eq
where
\bq\label{eq10.1}
f(p) = 3 p^{\sfrac{1}{6}} \sqrt{\bar F(p)}-\frac{9}{4}p^{-\sfrac{1}{3}} \bar F(p).
\eq

{Since the components} of the metric (\ref{eq10}) are polynomials {
in $t$, we may} look for the corresponding components of the embedding {function} $\tilde y^A(t,p)$ also in the form of polynomials of $t$ {similarly to the one used in the work \cite{statja27} for construction of the "cubic"{} embedding} of the Schwarzschild metric. The embedding was found for the case of 4-dimensional ambient space with the signature $(+-+-)$. Using the light cone coordinates in the ambient space $\tilde y^\pm=\tilde y^0\pm\tilde y^1$, it can be written in the form
\begin{align}\label{sp9}
	\tilde y^{+} &= 2t^3 + \frac{9}{8}t^2+tf(p), \nonumber\\
	\tilde y^2 &=  w(p), \\
	\tilde y^{-} &= t, \nonumber\\
	\tilde y^{3} &= \sqrt{\frac{3}{2}}t^2 -w(p)\nonumber,
\end{align}
where
\disn{sp9.1}{
w(p)=\frac{1}{2\sqrt{6}}\ls\int\!\sqrt{\bar F(p)}\,p^{-\sfrac{5}{6}} dp- f(p)\rs=\ns=
\frac{1}{\sqrt{6}}\te\ls p-\frac{9}{4}F_0\rs
\ls\ls\frac{9}{4}F_0\rs^{\sfrac{1}{2}}p^{\sfrac{1}{6}}+\frac{9}{8}F_0p^{-\sfrac{1}{6}}-\frac{3}{2}\ls\frac{9}{4}F_0\rs^{\sfrac{2}{3}}\rs
\nom}
and $\te(z)$ --- Heaviside step-function {(the explicit form \eqref{sp6} of the function $\bar F(p)$ was used to derive \eqref{sp9.1} ).}

Returning to the more natural coordinates $ \tau $, $ r $ and combining the components \eqref{sp9} with the remaining components \eqref{sp7} of the full embedding function $y^a(x^\mu)$, we finally get the embedding {function} for the metric (\ref{metric2}) in the form
\begin{align}\label{sp10}
	y^0 &= \tau^2 + \frac{9}{16}\tau^{\sfrac{4}{3}}+\frac{1}{2}\tau^{\sfrac{2}{3}} \ls f\ls\frac{r^3}{\tau^2}\rs+1\rs, \nonumber\\
	y^1 &= \tau^2 + \frac{9}{16}\tau^{\sfrac{4}{3}}+\frac{1}{2}\tau^{\sfrac{2}{3}} \ls f\ls\frac{r^3}{\tau^2}\rs-1\rs, \nonumber\\
	y^2 &=  w\ls\frac{r^3}{\tau^2}\rs, \\
	y^3 &= \sqrt{\frac{3}{2}}\tau^{\sfrac{4}{3}}-w\ls\frac{r^3}{\tau^2}\rs, \nonumber\\
	y^4 &= r \cos{\theta}, \nonumber\\
	y^5 &= r \sin{\theta} \cos{\varphi},  \nonumber\\
	y^6 &= r \sin{\theta} \sin{\varphi},\nonumber
\end{align}
the ambient space is 7-dimensional and has the signature $(+-+----)$.
Recall that the time $\tau$ is always assumed to be negative, and its zero value corresponds to the falling of all particles of the dust ball to the point $ r= 0$.

\begin{figure}[b]
\centering
\includegraphics[width=0.45\linewidth]{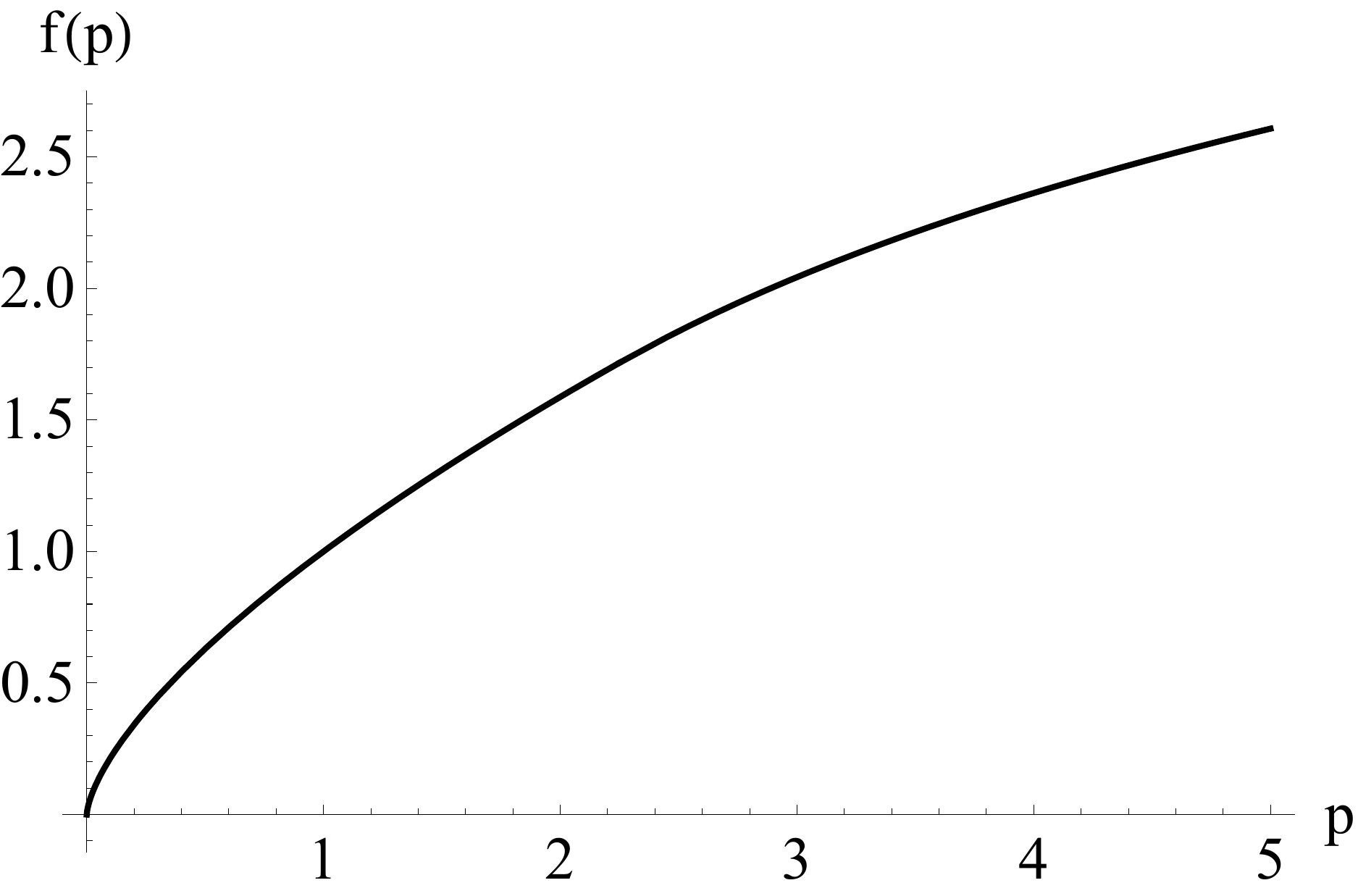}
\includegraphics[width=0.45\linewidth]{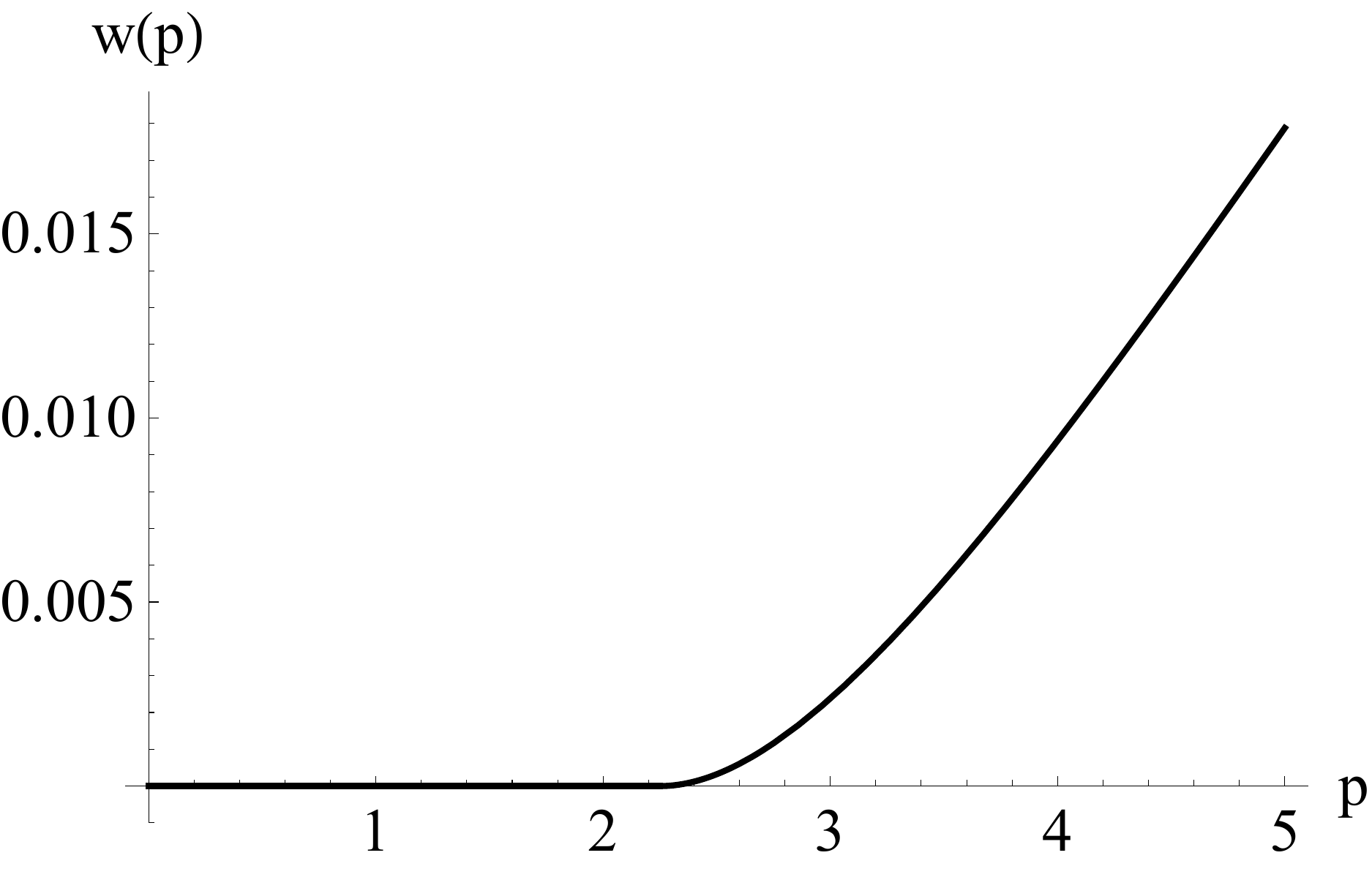}
\caption{\label{graf-f-w}
Graphs of the functions $f(p)$ and $w(p)$ {which {were used} in embedding {function} \ref{sp10}} with $F_0 = 1$.}
\end{figure}
The constructed embedding is global in the sense that it remains smooth for all values of $ r $ and all $ \tau <0 $, i.e. until the formation of a singularity.
Let's study the degree of smoothness for $ \tau <0 $.
First of all, when using spherical coordinates, it is necessary {to verify the smoothness} at the point $r=0$. {It is enough to check that the components $ y ^ 0 $, $ y ^ 1 $, $ y ^ 2 $, $ y ^ 3 $ at this point are finite and that their expansion in $ r $ do not contain odd degrees.} Since when $p<9F_0/4$ it turns out that $f(p)=p^{\sfrac{2}{3}}$, $w(p)=0$ (see \eqref{sp6},\eqref{eq10.1},\eqref{sp9.1}), the previous condition is satisfied.
Further, we should verify the smoothness on the boundary of the dust ball, which corresponds to {the argument} $p=9F_0/4$ of the functions $f(p)$, $w(p)$.
A simple analysis of the same formulas shows that at the indicated point the functions $f(p)$ and $w(p)$ are continuous together with their first derivatives, but their second derivatives have a jump. The graphs of these functions for $F_0=1$ are shown in Fig.~\ref{graf-f-w}.

\begin{figure}[b!]
\centering
\includegraphics[width=0.65\linewidth]{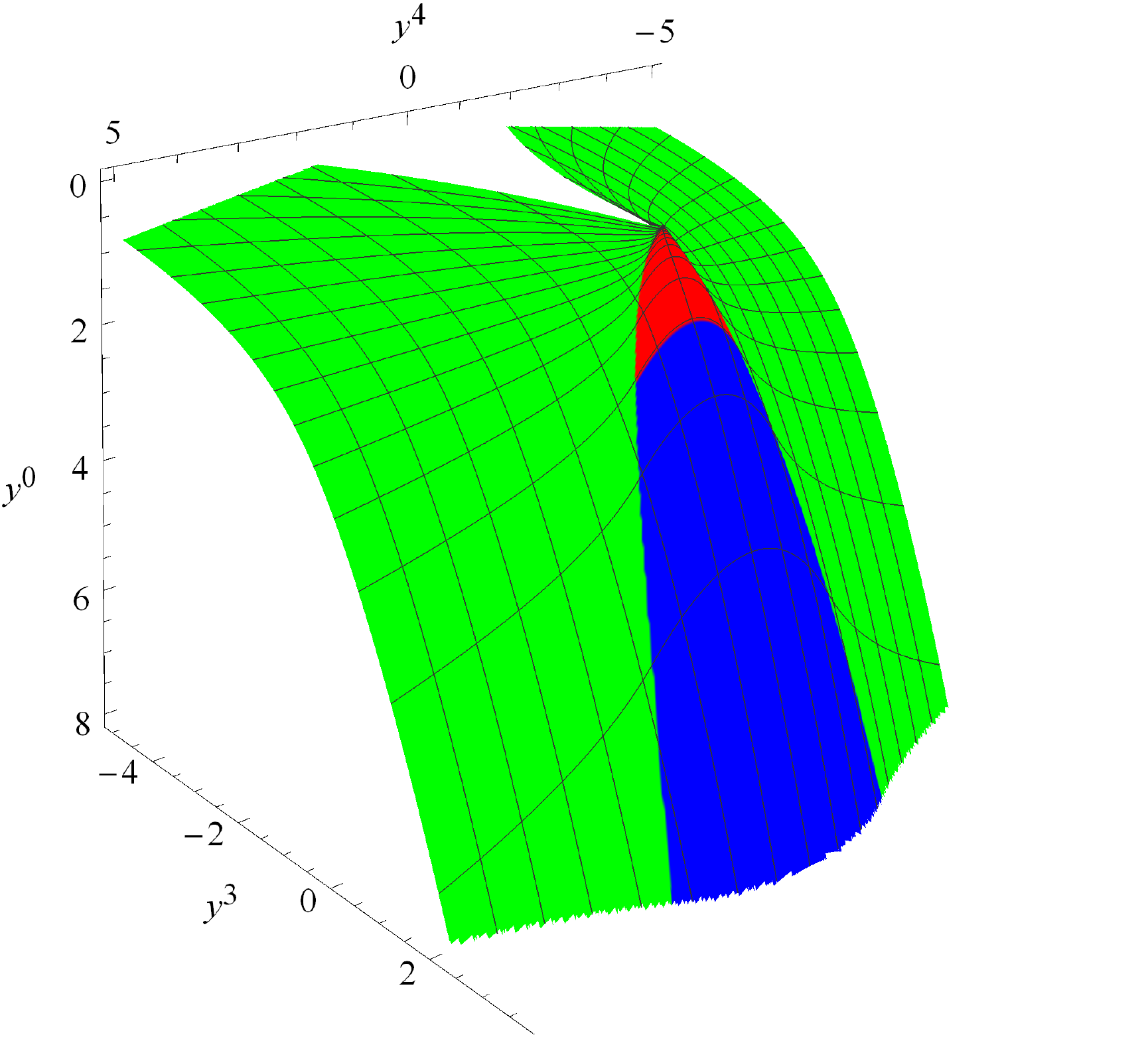}
\caption{\label{pic_emb1}
The projection of the two-dimensional submanifold $ \theta, \varphi = const $ of the embedding \eqref{sp10} onto the subspace $ y^0, y^3, y^4 $.
}
\end{figure}

{Such behavior of the functions $f(p)$ and $w(p)$ shows that} the constructed explicit embedding {functions \eqref{sp10} are} continuously differentiable.
The discontinuity of its second derivatives corresponds to the jump of the matter density on the boundary of the dust ball, so the smoothness of the embedding corresponds to the physical formulation of the problem. The projection of the two-dimensional surface corresponding to \eqref{sp10} with fixed angles $ \theta, \varphi $ onto the three-dimensional subspace $ y^0, y^3, y^4 $ is shown in Fig.~\ref{pic_emb1}.
Green
shows the area outside the dust ball,
blue
and
red
show the area of dust before and after horizon formation, respectively.
The singularity resulting from the collapse of the singularity corresponds to the value of $ y^0=0 $. It is interesting to note that in the neighborhood of this point the form of the surface turns out to be similar to the behavior near the singularity of the embedding \cite{robertson1933} for the spatially flat FRW model.

The variant of the embedding found in this section uses the ambient space with two time-like directions, which can be {considered as} its disadvantage {in attempts to give a physical meaning to the ambient space.}
In the next section we will try to eliminate this disadvantage by using another way of constructing an explicit form of embedding, based on matching the embeddings of two parts of the space-time corresponding to the areas inside and outside the dust ball.

\section{Constructing the embedding {by means of} a junction of solutions}
We will look for the embedding of the metric \eqref{metric}, written in the comoving coordinate frame, although in this framework the metric has coordinate singularity (see {the remark} before {Eq.} \eqref{sp1}).
We will {modify the well-known embeddings for the areas inside (FRW metric) and outside (Schwarzschild metric) of the dust ball,} so that the resulting embedding functions can be joined.

In comoving coordinates, $\chi$ labels each spherical shell of the dust cloud, therefore $\exists \chi_0>0$ such that $\chi=\chi_0$ marks the boundary of the dust ball.
At that, the area $0 \leqslant \chi < \chi_0$ contains matter and the area $\chi > \chi_0$ corresponds to the empty space.
We choose the function $r(\tau, \chi)$ in the {Eq.} (\ref{v123}) according to the third option, which corresponds to the closed FRW model {for the homogeneous matter density.}
If we choose the functions
\bq\label{vibor}
F(\chi) = \frac{R \sin^3{\chi}}{\sin^3{\chi_0}}, \qquad f(\chi) = -\sin^2{\chi}, \qquad \tau_0(\chi) = const,
\eq
we {obtain from \eqref{v123} the expression for $r(\tau, \chi)$ in the region $0 \leqslant \chi < \chi_0$ in the following form:}
\bq\label{sp14}
	r(\tau, \chi) = \frac{R \sin{\chi}}{2 \sin^3{\chi_0}}  E \left( \pi - \frac{2 \sin^3{\chi_0}}{R} \tau \right).
\eq
Then according to \eqref{EMT} we obtain
\begin{equation}
\rho(\tau, \chi) = \frac{24 \sin^6{\chi_0} }{\varkappa \, R^2 \, E^3 \left( \pi - \frac{2 \sin^3{\chi_0}}{R} \tau \right)},
\end{equation}
which show, that matter density doesn't depend on $\chi$. It follows that the choice \eqref{vibor} corresponds to the homogeneous matter distribution.
{Here, $R$ is the Schwarzschild radius of the whole dust ball} defined by its total mass, and the parameter $\chi_0$ determines the maximal size of the ball $r_{max} = R/\sin^2{\chi_0}$, since {according to Eq. \eqref{sp4},} the function $E(x)$ takes values from $0$ to $2$.
The value of $\tau_0$ is chosen so that the moment $\tau = 0$ corresponds to the maximal size of the ball.
The time $\tau$ varies in the finite limits $\tau \in [-\pi R / (2 \sin ^ 3 {\chi_0}), \pi R / (2 \sin^3{\chi_0})]$,
moreover, the initial and final values correspond to the emergence of matter from the singularity in the past and its fall into the singularity in the future.

A substitution of the function $r(\tau, \chi)$ {from Eq. \eqref{sp14} into Eq. (\ref{metric})} gives the FRW metric
\bq\label{sp19}
ds^2 = d\tau^2 - a^2(\tau) \left(d\chi^2 + \sin^2{\chi}d\Omega^2 \right)
\eq
with a scale {parameter:}
\bq\label{spp1}
a(\tau) = \frac{R}{2 \sin^3{\chi_0}} E \left( \pi - \frac{2 \sin^3{\chi_0}}{R} \tau \right).
\eq

{When $\chi > \chi_0$,} we can choose
{
\bq
	F(\chi) = R, \quad f(\chi) = -\frac{2}{r_m(\chi)}, \quad \tau_0(\chi) = \frac{ \pi r_m^{\sfrac{3}{2}}(\chi)}{2 R^{\sfrac{1}{2}}}
\eq
and {from} the formula \eqref{v123} we obtain}
\bq\label{spp1a}
r(\tau, \chi) = \frac{r_m(\chi)}{2} E \left( \pi - \frac{2R^{\sfrac{1}{2}}}{r^{\sfrac{3}{2}}_m(\chi)} \tau \right).
\eq
After substitution into \eqref{metric}, it gives the Schwarzschild metric in some coordinate frame
because $F$ is constant, see the text after equation \eqref{EMT} or \cite{novfrol} for more details.
The function $ r_m (\chi) $ {above} is only required to increase monotonically from the value of $ R / \sin^2{\chi_0} $ {at} $ \chi = \chi_0 $ to infinity {at} $ \chi \to \infty $. {Apart from this,} $r_m(\chi)$ can be chosen arbitrary.

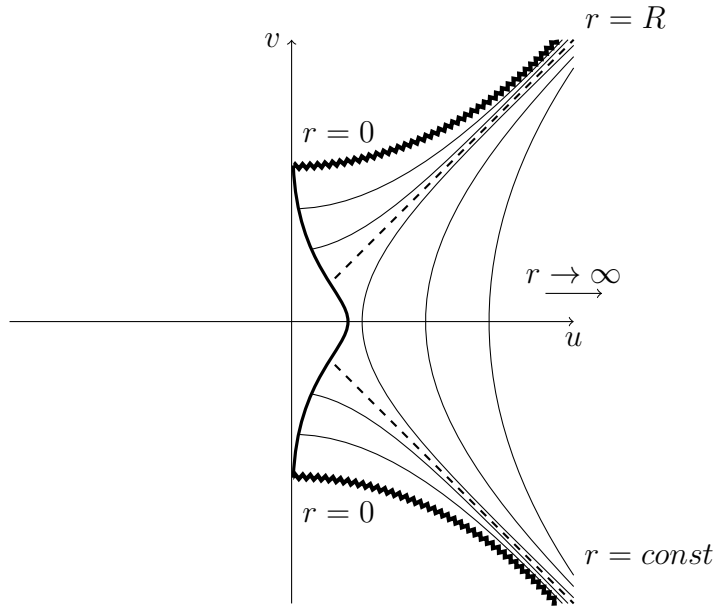
\begin{figure}[h!]
	\centering
		\begin{tikzpicture}[scale=0.75]
			\draw[->] (-5, 0) -- (5, 0)node[below]{$u$};
			\draw[->] (0, -5) -- (0, 5)node[left]{$v$};
			\draw[->] (4.5, 0.5) -- (5.5, 0.5)node[midway,above]{$r \to \infty$};
			\begin{scope}
			\clip (7.5, 6) -- (5, 6) -- (0, 3.5) .. controls (0, 1) and (1, 0.5) .. (1, 0) -- (1, 0) .. controls (1, -0.5) and (0, -1) .. (0, -3.5) -- (5, -6) -- (7.5, -6) -- cycle;
			\draw[thick, dashed] (-5, -5) -- (5, 5)node[above right]{$r = R$};
			\draw[thick, dashed] (-5, 5) -- (5, -5);
			\draw (5, 4.9) .. controls (0, 0) and (0, -0) .. (5, -4.9);
			\draw (5, 4.7) .. controls (1.5, 1) and (1.5, -1) .. (5, -4.7);
			\draw (5, 4.5) .. controls (3, 1.5) and (3, -1.5) .. (5, -4.5)node[above right]{$r=const$};
			\draw (-5, 4.9) .. controls (0, 0) and (0, -0) .. (-5, -4.9);
			\draw (-5, 4.7) .. controls (-1.5, 1) and (-1.5, -1) .. (-5, -4.7);
			\draw (-5, 4.5) .. controls (-3, 1.5) and (-3, -1.5) .. (-5, -4.5);
			\draw (4.9, 5) .. controls (0, 0) and (0, 0) .. (-4.9, 5);
			\draw (4.8, 5) .. controls (1, 1) and (-1, 1) .. (-4.8, 5);
			\draw[ultra thick, decoration = {zigzag,segment length = 1mm, amplitude = 0.3mm}, decorate] (4.7, 5) .. controls (2, 2) and (-2, 2) .. (-4.7, 5);
			\draw (0, 3)node[above right]{$r = 0$};
			\draw (4.9, -5) .. controls (0, 0) and (0, 0) .. (-4.9, -5);
			\draw (4.8, -5) .. controls (1, -1) and (-1, -1) .. (-4.8, -5);
			\draw[ultra thick, decoration = {zigzag,segment length = 1mm, amplitude = 0.3mm}, decorate] (4.7, -5) .. controls (2, -2) and (-2, -2) .. (-4.7, -5);
			\draw (0, -3)node[below right]{$r = 0$};
			\end{scope}
			\draw[very thick] (0.03, 2.77) .. controls (0.1, 1) and (1, 0.5) .. (1, 0) -- (1, 0) .. controls (1, -0.5) and (0.1, -1) .. (0.03, -2.77);
		\end{tikzpicture}
	\caption{\label{closedkrusk}The Kruskal diagram for the collapse of matter, which flew out of the white hole singularity.}
\end{figure}

The region $\chi > \chi_0$ is {presented} in the Kruskal diagram shown in Fig. \ref{closedkrusk}. It shows that in this case, the extreme and all internal particles of matter fly out of the white-hole singularity, reach the maximum distance
{placed outside of the horizon and then} collapse into a black-hole singularity. The rest of the manifold is described by the FRW geometry, as in the previous case.

\subsection{General case}
According to \cite{kasner2}, the minimal ambient space dimension for the Schwarzschild metric is~$6$. Therefore, the known five-dimensional embeddings for the FRW metric \cite{robertson1933, rosen65} should be modified by adding some components to the embedding {function}. The basic idea is that we should not change the dependence of the embedding {function} on the coordinate $\chi$, but only add components depending on $\tau$. In this approach, the condition for the fulfillment of the embedding equations \eqref{s1} is reduced to solving an ODE for the {components of embedding function}.

Hereinafter we will denote $y^a_f$ the embedding {function}, related to the FRW metric, and $y^a_s$ --- to the Schwarzschild metric.
The five-dimensional embedding of the metric of the closed FRW model has form
\begin{align}
\label{sp13}	y^0_f &= h(\tau), \\
\label{0}	y^1_f &= a(\tau) \cos{\chi}, \\
\label{1}	y^2_f &= a(\tau) \sin{\chi} \cos{\theta}, \\
\label{2}	y^3_f &= a(\tau) \sin{\chi} \sin{\theta} \cos{\varphi}, \\
\label{3}	y^4_f &= a(\tau) \sin{\chi} \sin{\theta} \sin{\varphi}
\end{align}
with signature $(+----)$, where $a(\tau)$ is defined by the formula \eqref{spp1}, and $h(\tau)$ should be found from the equation \eqref{s1}.
The component $y^0_f$ will be modified in the process of the junction, and the remaining  components we won't change. Then for $ \chi = \chi_0 $ this block of components must coincide with any four {components of embedding function} of the Schwarzschild metric embedding.

We will use the well-known global (that is, smoothly covering areas both outside and inside the horizon) 6-dimensional embeddings \cite{statja27} of the Schwarzschild metrics as the basis for the junction.
There are four such embeddings: Fronsdal embedding, Davidson-Paz embedding, asymptotically flat embedding and cubic with respect to time embedding, see details in \cite{statja27}.
The embedding function for all of them consists of the components $ \{ \tilde y_s^A(t, r), \hat y_s^i (r, \te, \ff) \} $ (here $ A = 0,1,2 $ ), three of which $ \hat y_s^i (r, \te, \ff) $ have the abovementioned form \eqref{sp7} and after substitution $ r = r (\tau, \chi) $ they will coincide on the boundary of the ball with (\ref{1}) - (\ref{3}).
But the function (\ref{0}) in the general case does not coincide with any of the known expressions for the remaining components $ \tilde y_s^A (t (\tau, \chi), r (\tau, \chi)) $, where the {function $t(\tau, \chi)$ is the} Schwarzschild time in the comoving coordinate frame {(see \cite{misner}):}
\disn{sp11}{
t(\tau, \chi) = R \ln{\left| \frac{\sqrt{\frac{r_m(\chi)}{R} -1}+\sign(\tau)\sqrt{\frac{r_m(\chi)}{r(\tau, \chi)} -1}}{\sqrt{\frac{r_m(\chi)}{R} -1 }-\sign(\tau)\sqrt{\frac{r_m(\chi)}{r(\tau, \chi)} -1 }}\right|} + \ns
+ R \sqrt{\frac{r_m(\chi)}{R} -1 } \left[\sign(\tau)\arccos{\left( \frac{2r(\tau, \chi)}{r_m(\chi)}-1\right)} + \frac{\tau}{\sqrt{R r_m(\chi)}} \right],
\nom}
where $\sign(\tau)=\pm1$ depending on the sign of $\tau$. Note that all expressions under the {root are} always non-negative.

For matching with {the component \eqref{0}}, we will artificially add
{to the embedding of the Schwarzschild metric new component
\disn{sp11nn}{
\tilde y_s^3 = r \ctg{\chi_0},
\nom}
expanding} it to seven-dimensional.
In addition, it is necessary to modify the block $ \tilde y^0_s, \tilde y^1_s, \tilde y^2_s $, {in order to avoid violation of the Eq. (\ref{s1}).}
This can be done for each of the four above-mentioned types of embeddings of the Schwarzschild metric since their construction reduces to solving an ODE {in} the variable $r$.
Adding {new} component $ \tilde y_s^3 $ {of simple form \eqref{sp11nn}} results in the appearance of the constant
term $\ctg^2{\chi_0}$ {in mentioned ODE} which does not violate the local solvability of {it}.

We {present} an explicit form of the result for the case when Fronsdal embedding \cite{frons} is taken as the basis.
Outside of the dust ball, i.e. when $\chi>\chi_0$, the embedding has the form:
\disn{fron}{
r(\tau, \chi)>R:\quad
y^0_s = w(\tau, \chi) \sh{\left( \frac{t(\tau, \chi)}{2 R}\right)}, \qquad
y^1_s = w(\tau, \chi) \ch{\left( \frac{t(\tau, \chi)}{2 R}\right)}, \no
r(\tau, \chi)<R:\quad
y^0_s = \sign(\tau) w(\tau, \chi) \ch{\left( \frac{t(\tau, \chi)}{2 R}\right)}, \quad
y^1_s = \sign(\tau) w(\tau, \chi) \sh{\left( \frac{t(\tau, \chi)}{2 R}\right)}, \no
y_s^2  = R \, q_{\chi_0} \left( \frac{r(\tau, \chi)}{R} \right), \qquad
y_s^3 = r(\tau, \chi) \ctg{\chi_0}, \no
y_s^4 = r(\tau, \chi) \cos{\theta}, \qquad y_s^5 = r(\tau, \chi) \sin{\theta} \cos{\varphi}, \qquad y_s^6 = r(\tau, \chi) \sin{\theta} \sin{\varphi},
\nom}
and inside, i.e. {when $\chi<\chi_0$:}
\disn{frid}{
r(\tau, \chi_0)>R:\quad
y^0_f = w(\tau, \chi_0) \sh{\left( \frac{t(\tau, \chi_0)}{2 R}\right)}, \qquad
y^1_f = w(\tau, \chi_0) \ch{\left( \frac{t(\tau, \chi_0)}{2 R}\right)}, \no
r(\tau, \chi_0)<R:\quad
y^0_f = \sign(\tau) w(\tau, \chi_0) \ch{\left( \frac{t(\tau, \chi_0)}{2 R}\right)}, \quad
y^1_f = \sign(\tau) w(\tau, \chi_0) \sh{\left( \frac{t(\tau, \chi_0)}{2 R}\right)}, \no
y_f^2 = R \, q_{\chi_0} \left( \frac{r(\tau, \chi_0)}{R} \right),\qquad
y_f^3 = a(\tau) \cos{\chi}, \no
y_f^4 = a(\tau) \sin{\chi} \cos{\theta}, \qquad y_f^5 = a(\tau) \sin{\chi} \sin{\theta} \cos{\varphi}, \qquad y_f^6 = a(\tau) \sin{\chi} \sin{\theta} \sin{\varphi},
\nom}
where
\bq\label{sp12}
w(\tau, \chi)=2 R \sqrt{\left|1-\frac{R}{r(\tau, \chi)}\right|},\qquad
q_{\chi_0}(x) = \int \limits_1^x d u \sqrt{\frac{1}{u^3}+\frac{1}{u^2}+\frac{1}{u}-\ctg^2{\chi_0}},
\eq
$a(\tau)$ is defined by {Eq.} \eqref{spp1}, and $r(\tau, \chi)$ --- {by} \eqref{spp1a}.
{It is clear that upon the modification of} the embedding of the FRW metric, {the component \eqref{sp13} } was replaced {by} the first three components of the embedding of the Schwarzschild metric taken at $\chi = \chi_0$, i.e. {they} are functions of $\tau$ {only}. Since the $g_{00}$ components of {metrics \eqref{metric} and \eqref{sp19}} are equal to $1$, the resulting set defines the FRW metric in the {area $ \chi <\chi_0 $.}

{Similarly to the case of Fronsdal embedding (see \cite{frons}), the embedding  defined by Eqs.~\eqref{fron} and \eqref{frid} with ambient space signature $(+------)$ is smooth at the horizon $r=R$.}
However, this embedding is not global, since it covers only the limited region of the values {of $r:$ for sufficiently large $ u $, the expression under the root in the integrand in \eqref{sp12} becomes negative.}

At the points $ \chi = \chi_0 $, the expressions \eqref{fron} and \eqref{frid} coincide, so the constructed embedding turns out to be continuous on the boundary of the dust ball.
However, it can be shown that the surface described by the embedding is not continuously differentiable: there is a kink at the joining boundary. {Thus,} for an arbitrary value of $ \chi_0 $ the embedding into a flat 7-dimensional space with one timelike direction, constructed in this section, is continuous, but {it is} not continuously differentiable, and {also, it is not global.}
It can be considered as some approximation to a global smooth embedding in the ambient space with a large number of dimensions.
In the next section, we will find a similar embedding with better properties by choosing a certain value of $ \chi_0 $.

\subsection{Special case $\chi_0 = \pi/2$}
\begin{figure}[h!]
\centering
\begin{tikzpicture}[scale=0.75]
	\draw[->] (-5, 0) -- (5, 0)node[below]{$u$};
	\draw[->] (0, -5) -- (0, 5)node[left]{$v$};
	\draw[->] (4.5, 0.5) -- (5.5, 0.5)node[midway,above]{$r \to \infty$};
	\begin{scope}
	\clip (7.5, 6) -- (5, 6) -- (0, 3.5) .. controls (0, 1) and (0, 0.5) .. (0, 0) -- (0, 0) .. controls (0, -0.5) and (0, -1) .. (0, -3.5) -- (5, -6) -- (7.5, -6) -- cycle;
	\draw[thick, dashed] (-5, -5) -- (5, 5)node[above right]{$r = R$};
	\draw[thick, dashed] (-5, 5) -- (5, -5);
	\draw (5, 4.9) .. controls (0, 0) and (0, -0) .. (5, -4.9);
	\draw (5, 4.7) .. controls (1.5, 1) and (1.5, -1) .. (5, -4.7);
	\draw (5, 4.5) .. controls (3, 1.5) and (3, -1.5) .. (5, -4.5)node[above right]{$r=const$};
	\draw (-5, 4.9) .. controls (0, 0) and (0, -0) .. (-5, -4.9);
	\draw (-5, 4.7) .. controls (-1.5, 1) and (-1.5, -1) .. (-5, -4.7);
	\draw (-5, 4.5) .. controls (-3, 1.5) and (-3, -1.5) .. (-5, -4.5);
	\draw (4.9, 5) .. controls (0, 0) and (0, 0) .. (-4.9, 5);
	\draw (4.8, 5) .. controls (1, 1) and (-1, 1) .. (-4.8, 5);
	\draw[ultra thick, decoration = {zigzag,segment length = 1mm, amplitude = 0.3mm}, decorate] (4.7, 5) .. controls (2, 2) and (-2, 2) .. (-4.7, 5);
	\draw (0, 3)node[above right]{$r = 0$};
	\draw (4.9, -5) .. controls (0, 0) and (0, 0) .. (-4.9, -5);
	\draw (4.8, -5) .. controls (1, -1) and (-1, -1) .. (-4.8, -5);
	\draw[ultra thick, decoration = {zigzag,segment length = 1mm, amplitude = 0.3mm}, decorate] (4.7, -5) .. controls (2, -2) and (-2, -2) .. (-4.7, -5);
	\draw (0, -3)node[below right]{$r = 0$};
	\end{scope}
	\draw[ultra thick] (0, 2.77) .. controls (0, 1) and (0, 0.5) .. (0, 0) -- (0, 0) .. controls (0, -0.5) and (0, -1) .. (0, -2.77);
\end{tikzpicture}
\caption{\label{speckrusk}The Kruskal diagram for the special case {when the} matter does not leave the limits of the Schwarzschild radius.}
\end{figure}
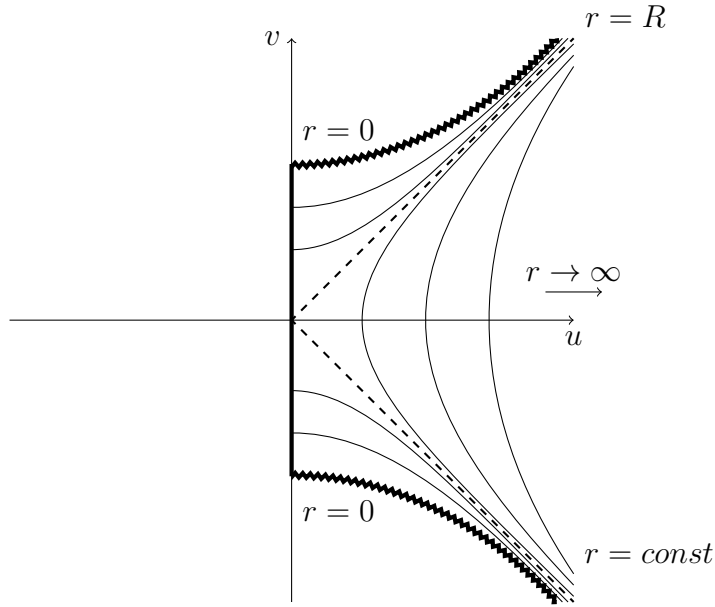

If $\chi_0 = \pi/2${, the} maximal radius of the ball $r_{max} = R/\sin^2{\chi_0}$ (see \eqref{sp14}) is equal to the Schwarzschild radius $R$.
This means that in the process of its movement, dust matter does not go out from under the horizon, {and therefore, the} junction will always occur at values of $ r \leqslant R $. The corresponding Kruskal diagram is shown in Fig.~\ref{speckrusk}.

In the case of $ \chi_0 = \pi/2 $, the embedding \eqref{fron} extended to 7-dimensional {one} becomes 6-dimensional again (since $ y^3_s $ turns out to be identically zero) and reduces to the original Fronsdal's embedding \cite{frons}.
{For $ \chi_0 = \pi/2 $, when $ r_m(\chi_0) = R $, the function $ t (\tau, \chi_0) = 0 $, according to \eqref{sp11}, and therefore, in \eqref{fron} the component $ y_s^1 $ vanishes on the {junction} boundary for all $ \tau $.}
{Since the same is true for the function (\ref{0}) in the embedding of the FRW metric, just these components can be identified.}
As a {result, there is no need to extend the Fronsdal embedding to a seven-dimensional one (as was done in the previous section), and in the embedding of the FRW metric it is enough to replace the component \eqref{sp13} by the set of two components $y^0_s$, $y^2_s$ from} \eqref{fron} ({when $r(\tau, \chi_0)<R$}) taken at $\chi=\chi_0=\pi/2$.
{Thus,} we obtain the following embedding: outside of the dust ball (where $\chi>\chi_0$) it is defined by the components $y^0_s,y^1_s,y^2_s,y^4_s,y^5_s,y^6_s$, {given by} \eqref{fron}, and inside (where $\chi<\chi_0$) it is defined by the components $y^0_f$ ({for} $r(\tau, \chi_0)\le R$), $y^2_f,y^4_f,y^5_f,y^6_f$, {given by} \eqref{frid}, {together with} the component $y^1_f$, {defined by} \eqref{0} {but with the} opposite sign.

{As mentioned above,} in the coordinates of $ \tau, \chi $ the metric (\ref{metric}) has a coordinate singularity.
{Therefore, this coordinates are inconvenient to express the resulting embedding and
we will use the coordinates of the ambient space $y^0$ and $y^1$ instead of them.}
{Taking into account} that the junction boundary corresponds to $y^1 = 0$,
{{it is easy to see that for embedding described above} the part of embedding outside the dust ball corresponds to $ y^1> 0$,
and the part of embedding inside the dust ball corresponds to $ y^1< 0 $. {As a result,} for $ y^1>0$, i.e. outside the dust ball,
we can write (expressing the value $ r(\tau, \chi) $ through $ y^0, y^1$) remaining components of this embedding:}
\begin{align}\label{sp16}
y^2 &= R \, q_{\chi_0} \left( \frac{\tilde r(y^0,y^1)}{R} \right), \nonumber\\
y^3 &= \tilde r(y^0,y^1) \cos{\theta}, \\
y^4 &= \tilde r(y^0,y^1) \sin{\theta} \cos{\varphi},\nonumber\\
y^5 &= \tilde r(y^0,y^1) \sin{\theta} \sin{\varphi}, \nonumber
\end{align}
where
\disn{sp17}{
\tilde r(y^0,y^1)=\frac{4R^3}{{y^0}^2-{y^1}^2+4R^2}.
\nom}
{For $y^1<0$, i.e. inside the dust ball, we can write
(expressing value $r(\tau, \chi_0)=a(\tau)$ through $y^0,y^1$):}
\begin{align}\label{sp18}
y^2 &= R \, q_{\chi_0} \left( \frac{\tilde r(y^0,0)}{R} \right),\nonumber\\
y^3 &= \sqrt{\tilde r(y^0,0)^2-{y^1}^2} \cos{\theta}, \\
y^4 &= \sqrt{\tilde r(y^0,0)^2-{y^1}^2} \sin{\theta}\cos{\varphi},\nonumber\\
y^5 &= \sqrt{\tilde r(y^0,0)^2-{y^1}^2} \sin{\theta}\sin{\varphi}. \nonumber
\end{align}

The resulting embedding in the space with the signature $ (+-----) $ is global, covering all values of $ r $ (note that for $ \chi_0 = \pi/2 $ the function $ q_{\chi_0}(x) $ remains real for all values of its argument) and {the whole interval of time from the moment when matter} leaves the singularity in the past until it falls into the singularity in the future.
At the same time, it turns out to be smooth both on the horizon $ r = R $, which in this case is completely outside the dust ball, and on the boundary of the dust ball $ y ^ 1 = 0 $, where embedding {function \eqref{sp16}, \eqref{sp18} is} not only continuous but also continuously {differentiable}
{(this is easy to check {comparing} the values of derivatives of the components \eqref{sp16} and \eqref{sp18} with respect to $y^1$ at the point $y^1=0$).}
{Meanwhile, their second derivatives have} a jump.

\begin{figure}[h!]
\centering
\includegraphics[width=0.55\linewidth]{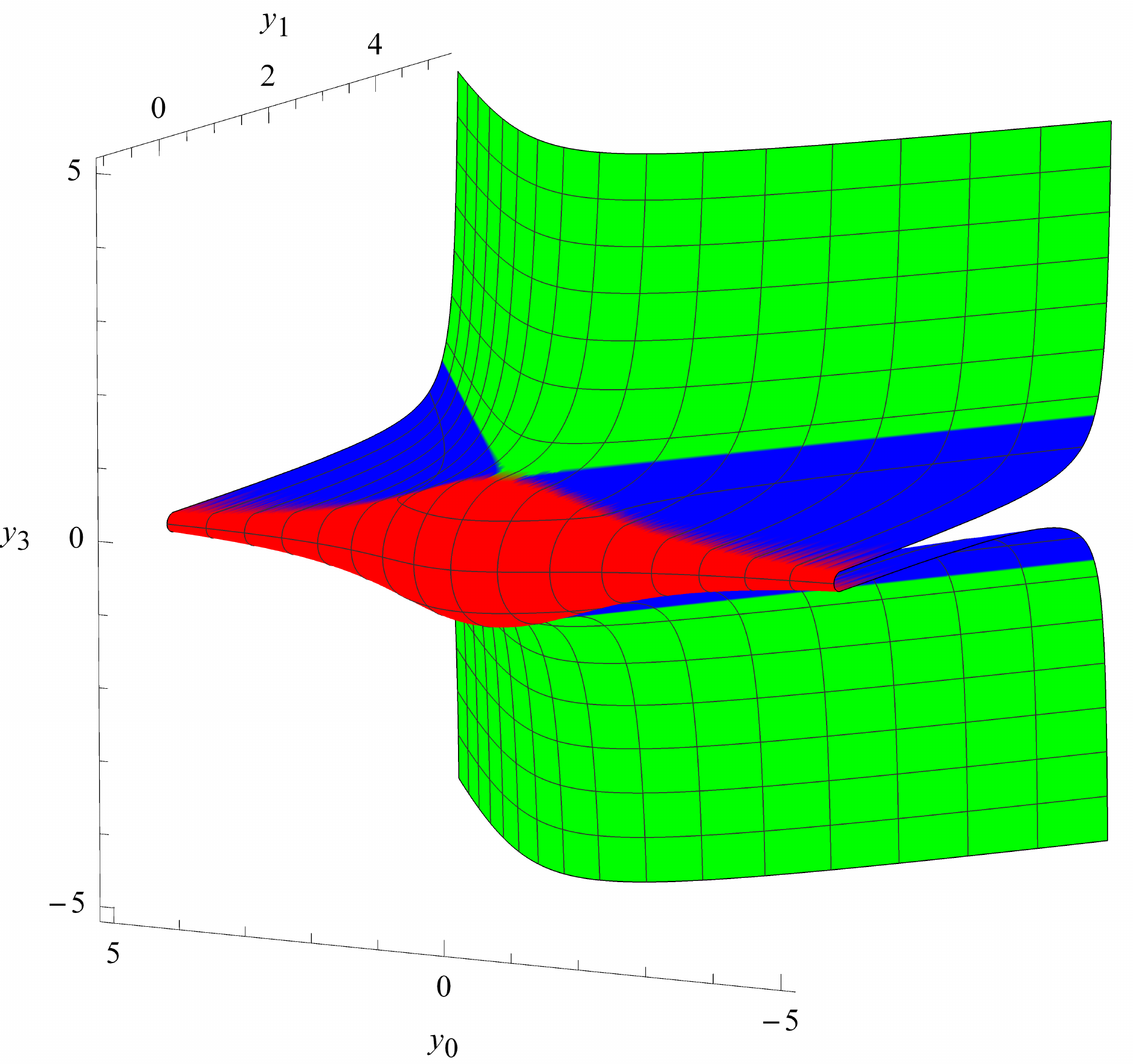}
\caption{\label{pic_emb}
The projection of the two-dimensional submanifold $ \theta, \varphi = const $ of the embedding \eqref{sp16}, \eqref{sp18} onto the subspace $ y^0, y^1, y^3 $.
}
\end{figure}

The projection of the two-dimensional {surface \eqref{sp16}, \eqref{sp18},} with fixed angles $ \theta, \varphi, $ onto the three-dimensional subspace $ y^0, y^1, y^3 $ is shown in Fig.~\ref{pic_emb}.
Red
shows the area of dust,
green
and
blue
show the area outside the dust ball above and below the horizon, respectively.
The singularities in the past and the future correspond to the limits of $ y^0 \to \pm \infty $.

The constructed embedding \eqref{sp16}, \eqref{sp18}, as well as the embedding \eqref{sp10} obtained in Section~3, has smoothness corresponding to the physical formulation of the problem (a jump in the density of matter).
{The advantage of embedding \eqref{sp16}, \eqref{sp18} is the presence of only one time-like direction in the ambient space. However, it describes a physically less interesting {situation} of motion of matter under the eternally existing horizon, while embedding \eqref{sp10} describes the {dynamical} formation of the horizon.}

{\bf Acknowledgements}
The work of A.~K. is supported by RFBR Grant No.~18-31-00169, the work of S.~P. is supported by RFBR Grant No.~20-01-00081.



\end{document}